# About the aftershocks, the Omori law, and the Utsu formula

Anatol Guglielmi[1], Alexey Zavyalov[1], Oleg Zotov[1,2], and Boris Klain[2]

[1] *Schmidt Institute of Physics of the Earth, Russian Academy of Sciences; Bol'shaya Gruzinskaya str., 10, bld. 5 1, Moscow, 123242 Russia; guglielmi@mail.ru (A.G.), zavyalov@ifz.ru (A.Z.); ozotov@inbox.ru (O.Z.)*

[2] *Borok Geophysical Observatory of Schmidt Institute of Physics of the Earth, Russian Academy of Sciences; klb314@mail.ru (B.K.)*

**Abstract:** After the main shock of an earthquake, a stream of aftershocks that does not subside for a long time is usually observed. Fusakichi Omori found that the frequency of aftershocks decreases hyperbolically with time. It has recently been observed that Omori's law can be viewed as a solution to a differential equation describing the evolution of aftershocks. An alternative way of describing is based on Utsu law, which states that the frequency of aftershocks decreases with time according to a power law. The presented paper is polemical. We discuss the issue of the applicability of each of the three alternative ways of describing aftershocks. The Omori law has a limited scope. The law is valid only in the so-called Omori epoch, after which the earthquake source undergoes a bifurcation. In the Omori epoch, the Utsu law is also valid, but it does not differ in this epoch from the Omori law. The general conclusion is that the existence of the Omori epoch and the phenomenon of bifurcation exclude the possibility of describing by a continuous smooth function. At the same time, the differential evolution equation is applicable both before and after the bifurcation point.

## 1. Introduction

This year marks 100 years since the death of Fusakichi Omori [1–4]. He made an outstanding contribution to the physics of earthquakes. In 1894, while still a young man, he discovered that after a strong earthquake, the aftershock frequency $n(t)$ decreases on average hyperbolically with time:

$$n(t) = \frac{k}{c+t}. \qquad (1)$$

Here $k > 0$, $c > 0$, and $t \geq 0$ [5]. Not long ago [6], it was noticed that the Omori formula (1) is the general solution of the nonlinear differential equation

$$\frac{dn}{dt} + \sigma n^2 = 0. \qquad (2)$$

Here $\sigma$ is the deactivation coefficient of the source cooling down after the main shock of the earthquake. We see that the evolution equation (2) has a unique solution

$$n(t) = \frac{n_0}{1 + n_0 t}, \qquad (3)$$

which coincides with the Omori formula (1) up to notation. Thus, formula (1) and equation (2) are equivalent methods for describing the evolution of aftershocks.

Another way of describing aftershocks is widespread in the geophysical literature. Namely, instead of the Omori formula (1), a function

$$n(t) = \frac{k}{(c+t)^p} \qquad (4)$$

is used in which $p$ is a dimensionless additional parameter [7–9]. Formula (4) is often called the Utsu law (see, for example, [10] and references therein).

This paper is a polemical one. We will discuss a rather subtle issue regarding the scope of applicability of each of the three laws of aftershock evolution, (1), (2) and (4). In Section 2, we indicate the range of applicability of the hyperbolic law (1). In section 3, we describe the bifurcation phenomenon revealed using equation (2) and show that the bifurcation point limits the applicability region of (1). We present arguments in favor of the idea that formula (4) is inapplicable for describing aftershocks.

## 2. Omori epoch

So, we consider the source as a dynamic system. Our phenomenological theory (2) contains a parameter $\sigma$ that characterizes the system, and this parameter can, generally speaking, depend on time. The $\sigma(t)$ dependence reflects the non-stationarity of the geological environment in the source. Non-stationarity arises due to the fact that after formation of the main rupture, the transition of rocks from one state of quasi-equilibrium to another begins. In other words, the source relaxes, which manifests itself in a series of aftershocks.

Let us introduce the so-called proper time of the source

$$\tau = \int_0^t \sigma(t')dt' . \tag{5}$$

The master equation will take the form

$$\frac{dn}{d\tau} + n^2 = 0 . \tag{6}$$

The general solution of equation has the form

$$n(\tau) = \frac{n_0}{1 + n_0 \tau} . \tag{7}$$

Solution (7) preserves the hyperbolic structure of Omori's law (1), with the difference that time in a non-stationary source flows unevenly.

Let us pose the inverse problem of source physics: Calculate the deactivation coefficient from experimental data on the frequency of aftershocks. We introduce an auxiliary function $g(t) = 1/n(t)$. The solution of the inverse problem is

$$\sigma = \frac{d}{dt}\langle g \rangle , \tag{8}$$

where the angle brackets denote the operation of smoothing the auxiliary function [11].

The practical solution of the inverse problem made it possible to reveal the existence of the Omori epoch, during which the source deactivation coefficient remains unchanged [12–15]. The duration of the Omori epoch varies from case to case from a few days to several months. Proper time in the Omori epoch is proportional to world time: $\tau = \sigma t$.

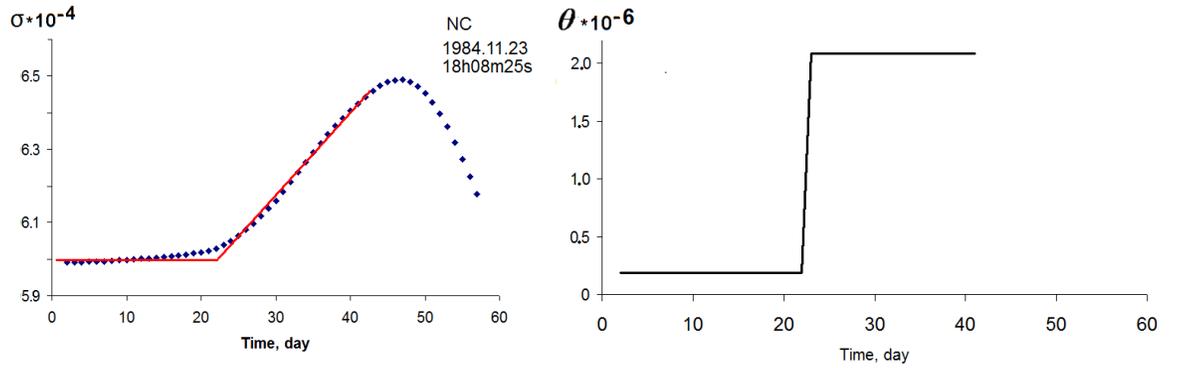

An example of solving the inverse problem is shown on the left. The event occurred in Northern California on November 23, 1984. The magnitude of the main shock M = 6. The dotted curve shows the variation in the deactivation factor. The solid broken line shows the fitting function. On the right, a jump in the time derivative of the deactivation coefficient is shown schematically. The dimension of $\theta$, plotted along the vertical axis, is $[\theta] = 1/\text{day}$.

Consider the figure [14, 15]. We see that evolution begins from the Omori epoch, with $\sigma = \text{const}$ over 20 days. At the end of the Omori epoch, there is a jump in the value of $\theta = d\sigma/dt$.

## 3. Discussion

The existence of the Omori epoch indicates that the Omori law (1) is applicable to describe the evolution of aftershocks. Its applicability, however, is limited in time. At the end of the epoch of Omori, there is a jump in the time derivative of the function $\sigma(t)$. Neither the hyperbolic law (1) nor the Utsu power law (4) describe the entire evolution of aftershocks as an integral process. Utsu law is applicable in the Omori epoch, but at this stage of evolution it does not differ from Omori law.

In [4, 16], the inverse problem was solved for eight aftershock series. Based on the analysis of solutions, the authors formulated a hypothesis that at the end of the Omori epoch, a bifurcation of the earthquake source occurs. The bifurcation phenomenon is well known in the theory of critical phenomena, in the theory of phase transitions, and in the theory of catastrophes [17–19]. Judging by the right side of the figure, we may be dealing with a first-order phase transition. However, it is difficult to talk about this at this stage of the study. We have yet to understand the mechanism of bifurcation within the framework of one or another model of the source, considered as a dynamic system.

### 4. Conclusion

The general conclusion is that the existence of the Omori epoch and the phenomenon of source bifurcation exclude the possibility of describing the evolution of aftershocks by a continuous smooth function. Hence it follows that neither the Omori law nor the Utsu formula can be used to describe the aftershock flow as an integral process. Omori's law is applicable only in the Omori epoch. In this epoch, the Utsu power function is also applicable, but in this epoch it coincides with Omori hyperbolic formula. At the end of the Omori epoch, the time derivative of the deactivation coefficient experiences a jump (rupture), after which another phase of aftershock evolution begins. At the same time, the nonlinear differential equation (2) describes evolution both in the Omori epoch and after it.

**Acknowledgments:** We are deeply grateful to A.L. Buchachenko for his attention to this study and for his valuable comments. We are grateful to the compilers of the earthquakes catalogs of Southern (https://scedc.caltech.edu) and Northern (http://www.ncedc.org) California, data from which were used in our study. The work was carried out according to the plan of state assignments of IPE RAS.

### References

1. *Davison, Ch.* Fusakichi Omori and his work on earthquakes // Bulletin of the Seismological Society of America. December 01, 1924. V. 14. No. 4. P. 240-255. DOI: https://doi.org/10.1785/BSSA0140040240 316

2. *Guglielmi A.V., Zavyalov A.D.* The 150$^{th}$ Anniversary of Fusakichi Omori (1868-1923) // IASPEI Newsletter, September 2018, p. 5-6.

3. *Guglielmi A., Klain B., Zavyalov A., Zotov O.* Omori Law. To the 100th anniversary of death of the famous Japanese seismologist // arXiv:2302.01277 [physics.geo-ph].

4. *Guglielmi A,, Zavyalov A., Zotov O, Klain B.* Omori Epoch: To the 100$^{th}$ anniversary of death of the famous Japanese seismologist
   // Applied Sciences. 2023 (presented).

5. *Omori F.* On the aftershocks of earthquake, J. Coll. Sci. Imp. Univ. Tokyo. 1894. V. 7. P. 111–200.

6. *Guglielmi A.V.* Interpretation of the Omori law // arXiv:1604.07017 [physics.geo-ph] // Izv., Phys. Solid Earth. 2016. V. 52. No. 5. P. 785–786. doi:10.1134/S1069351316050165.

7. *Hirano R.* Investigation of aftershocks of the great Kanto earthquake at Kumagaya // Kishoshushi. Ser. 2. 1924. V. 2. P. 77–83.

8. *Utsu T.* A statistical study on the occurrence of aftershocks // Geophys. Mag. 1961. V. 30. P. 521−605.

9. *Utsu T., Ogata Y., Matsu'ura R.S.* The centenary of the Omori formula for a decay law of aftershock activity // J. Phys. Earth. 1995. V. 43. № 1. P. 1–33.

10. *Salinas-Martínez A., Perez-Oregon J., AnaMaría Aguilar-Molina A., Muñoz-Diosdado A., Angulo-Brown F.* On the Possibility of Reproducing Utsu's Law for Earthquakes with a Spring-Block SOC Model // Entropy. 2023. V. 25, 816. https://doi.org/10.3390/e25050816.

11. *Guglielmi A.V.* Omori's law: a note on the history of geophysics // Phys. Usp. 2017. V. 60. P. 319–324. DOI: 10.3367/UFNe.2017.01.038039.

12. *Guglielmi A.V., Zavyalov A.D., Zotov O.D.*, A project for an Atlas of Aftershocks following large earthquakes // J. Volcanol. Seismol. 2019. V. 13. No. 6,. P. 415–419.

13. *Zavyalov A.D., Guglielmi A.V., Zotov O.D.* Three problems in aftershock physics // Journal of Volcanology and Seismology. 2020. V. 14. No. 5. P. 341–352.

14. *Guglielmi A.V., Klain B.I. Zavyalov A.D., Zotov O.D.* A Phenomenological theory of aftershocks following a large earthquake // J. Volcanology and Seismology. 2021. V. 15. No. 6. P. 373–378.

15. *Zavyalov, A., Zotov, O., Guglielmi, A., Klain, B.* On the Omori Law in the physics of earthquakes // Appl. Sci. 2022, vol. 12, issue 19, 9965. https://doi.org/10.3390/app12199965

16. *Guglielmi A., Zotov O.* Bifurcation of the earthquake source at the end of the Omori epoch // arXiv:2303.02582 [physics.geo-ph].

17. *Landau L.D., Lifshitz E.M.* Statistical Physics // Pergamon Press. 1970.

18. *Gilmore R.* Catastrophe Theory for Scientists and Engineers // Dover Publications. 1993.

19. *Guglielmi A.V.* Foreshocks and aftershocks of strong earthquakes in the light of catastrophe theory // Physics – Uspekhi. 2015. V. 58 (4). P. 384–397.